\pdfoutput=1

\documentclass[11pt]{article}

\usepackage[]{acl}

\usepackage{times}
\usepackage{latexsym}
\usepackage{graphicx}
\usepackage{booktabs}
\usepackage{pifont}
\usepackage{amsmath}
\usepackage{color}
\usepackage{bm}
\usepackage{amsfonts}
\usepackage{multirow}
\usepackage{bbding}
\usepackage{enumitem}
\usepackage{subcaption}
\usepackage{tcolorbox}
\usepackage{longtable}

\usepackage[T1]{fontenc}

\usepackage[utf8]{inputenc}

\usepackage{microtype}

\usepackage{inconsolata}

\newcommand{\dataname}[0]{LEAD}

\title{Enhancing Legal Case Retrieval via Scaling High-quality \\ Synthetic Query-Candidate Pairs}

\author{
Cheng Gao$^{2,1}$\thanks{Equal contribution.},
Chaojun Xiao$^{2,1*}$, Zhenghao Liu$^3$, \\ \textbf{Huimin Chen$^{4}$\thanks{Corresponding authors.}, 
Zhiyuan Liu$^1$, Maosong Sun$^{1\dagger}$}\\ 
    $^1$NLP Group, DCST, IAI, BNRIST, Tsinghua University, Beijing 
      $^2$Quan Cheng Laboratory \\
      $^3$Northeastern University 
      $^4$School of Journalism and Communication, Tsinghua University \\
 \texttt{\{gaoc24,xiaocj20\}@mails.tsinghua.edu.cn}, \texttt{\{huimchen,sms\}@tsinghua.edu.cn}
}

\begin{document}
\maketitle
\begin{abstract}
Legal case retrieval (LCR) aims to provide similar cases as references for a given fact description. This task is crucial for promoting consistent judgments in similar cases, effectively enhancing judicial fairness and improving work efficiency for judges. However, existing works face two main challenges for real-world applications: existing works mainly focus on case-to-case retrieval using lengthy queries, which does not match real-world scenarios; and the limited data scale, with current datasets containing only hundreds of queries, is insufficient to satisfy the training requirements of existing data-hungry neural models. To address these issues, we introduce an automated method to construct synthetic query-candidate pairs and build the largest LCR dataset to date, \dataname{}, which is hundreds of times larger than existing datasets. This data construction method can provide ample training signals for LCR models. Experimental results demonstrate that model training with our constructed data can achieve state-of-the-art results on two widely-used LCR benchmarks. Besides, the construction method can also be applied to civil cases and achieve promising results. The data and codes can be found in \url{https://github.com/thunlp/LEAD}.
\end{abstract}

\section{Introduction}
Legal case retrieval (LCR) aims to search for historically relevant cases based on a given fact description~\cite{DBLP:journals/ail/Bench-CaponAAABBBBCFGGLLLMPSSTTVWW12,DBLP:journals/ipm/BhattacharyaGPG22,DBLP:journals/corr/abs-2202-07209,DBLP:conf/sigir/YuS0DCXW22,DBLP:journals/is/SansoneS22}. This task can help legal professionals, such as judges and lawyers, improve work efficiency by providing past cases as references for current judgments. Thus, it plays a crucial role in promoting judicial fairness by facilitating similar cases receiving similar judgments.

\begin{figure}[t]
    \centering
    \includegraphics[width=\linewidth]{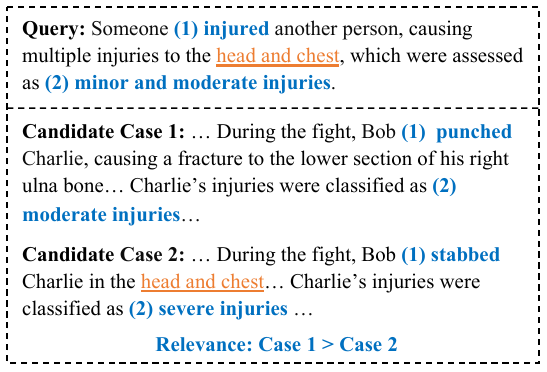}
    \caption{An example for legal case retrieval, where the key facts are in blue.}
    \label{fig:case}
\end{figure}

Different from open-domain retrieval, LCR demands a complex understanding of case details and necessitates models equipped with legal knowledge to generate knowledge-rich case representations~\cite{xiao2023legal,DBLP:conf/sigir/SunXZDW23}. As shown in Figure~\ref{fig:case}, models are required to recognize that the severity of injury rather than the location of injury is the key factor in assessing the relevance of given candidates to the query.
Recent years have seen significant efforts by scholars to improve the performance of LCR, including introducing additional knowledge features~\cite{DBLP:journals/ipm/BhattacharyaGPG22,leven,DBLP:conf/sigir/SunXZDW23} and designing LCR-oriented pre-training objectives~\cite{SAILER,CaseEncoder}.

However, despite these advancements, the real-world application of LCR still faces the following challenges: (1) \textbf{Asymmetric Retrieval.} Existing methods mostly focus on symmetric retrieval settings with lengthy fact descriptions for both queries and candidates. In contrast, real-world user queries often consist of only a few sentences describing key details. This inconsistency between application and training scenarios results in sub-optimal performance.
(2) \textbf{Limited Data.} Another challenge is the limited data scale, as legal data annotation requires highly skilled and experienced annotators, making it time-consuming and labor-intensive. Existing LCR datasets contain only a few hundred queries~\cite{lecard, lecardv2}, compared to tens of thousands in open-domain retrieval datasets~\cite{mmarco, Dureader, T2Ranking}. Besides, most retrieval methods rely heavily on data-hungry neural models, making the construction of large-scale, high-quality legal retrieval data a key to enhancing LCR performance.

To address these issues, this paper proposes a method for automatically constructing high-quality, synthetic legal retrieval datasets for model training.  
Specifically, given a case candidate, we employ a large-scale generative language model to first extract key facts, and omit entities, including names and places.
Then, based on the anonymous key fact, we require the model to generate a brief and coherent description of the case, which is regarded as the search query. In this way, the generated query is short and contains only a few sentences.
Additionally, to improve data diversity and enable the model to retrieve relevant cases even when key facts are not entirely consistent, we employ a knowledge-driven data augmentation strategy. For each query, we select the case that is most similar from the perspective of charges, related legal articles, and prison term, from the entire corpus as the augmented positive candidate.

This approach enables us to rapidly build the largest LCR dataset, \dataname{}, with over 100K query-candidate pairs and without any manual annotation, surpassing existing LCR datasets by a hundredfold. 
To verify the effectiveness of our method, we train dense passage retrieval models with \dataname{} and compare the model with several competitive baseline models, on two widely-used criminal LCR benchmarks. The experimental results demonstrate that models trained with our enriched high-quality case retrieval data can achieve state-of-the-art performance in LCR tasks. Besides, the proposed framework for data generation can be easily applied to civil case retrieval, and achieve satisfying performance. The code and data in our paper will be released to promote the development of LCR.

\section{Related Work}
\label{sec: Related Work}
\textbf{Legal Case Retrieval.}
Legal case retrieval is a challenging task that requires a deep understanding of legal documents. The task entails models identifying the most legally relevant cases within candidate documents concerning a given query case. 

Earliest work for LCR attempt to employ traditional retrieval models, including, BM25~\cite{BM25} and TF-IDF~\cite{TF-IDF}, for legal retrieval~\cite{DBLP:journals/ijlit/ZengWZK07}.
With the development of deep learning, many efforts have been devoted into designing neural architectures to enhance long textual representation~\cite{Longformer,DBLP:conf/ijcai/ShaoMLMSZM20}, interpretability~\cite{DBLP:conf/sigir/YuS0DCXW22,sun2023explainable}, legal knowledge enriched representation~\cite{DBLP:conf/ecir/AbolghasemiVA22,DBLP:journals/tois/MaWALSZM24,xiao2023legal,DBLP:journals/corr/abs-2210-11012,leven}.
Due to the lack of a large-scale LCR dataset, these researches mainly focus on the re-ranking phrase, overlooking the significance of dense passage retrieval (DPR) for high recall rate~\cite{DPR}. To elevate the data scarcity issues, some researchers explore the self-supervised pre-training for legal DPR. For instance, SAILER~\cite{SAILER} adopts an asymmetric encoder-decoder architecture, integrating various pre-training objectives to encode rich semantic information across tasks. CaseEncoder~\cite{CaseEncoder} leverages fine-grained legal provisions to select relevant and irrelevant cases for each query, thus improving the quality of training data. In this paper, we find that our data construction methods can further facilitate the LCR performance by scaling the high-quality instances for LCR.

\begin{figure*}
    \centering
    \includegraphics[width=0.9\textwidth]{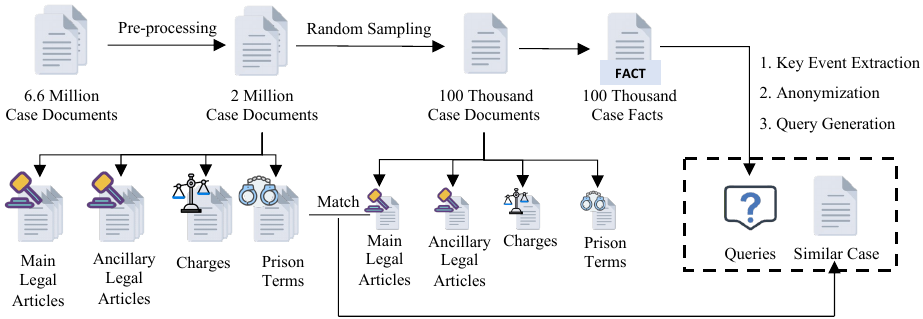}
    \caption{The illustration of the data construction process.}
    \label{fig:framework}
\end{figure*}

\textbf{Dataset for LCR.}
High-quality data lies in the core of existing data-hungry neural models for LCR. However, due to the highly skilled and experienced annotators required for legal data annotation, existing LCR datasets only contain a few hundred queries. For example, LeCaRD~\cite{lecard} consists of a total of 107 queries, each with 100 candidate documents, but only 30 of these documents have been manually annotated for relevance. LeCaRDv2~\cite{lecardv2} contains 800 queries, with only 30 documents per query annotated for relevance. CAIL2022-LCR is the competition dataset of the Challenge of AI in Law (CAIL). 
Compared to these datasets, open-domain retrieval datasets have hundreds of times more queries, such as T\textsuperscript{2}Ranking~\cite{T2Ranking} with 307k queries, and mMarco-Chinese~\cite{mmarco} with 516k queries. The lack of large-scale data hinders the development of LCR.

\textbf{Data Augmentation for Information Retrieval}
Data augmentation aims to increase the amount of training data by heuristically generating new data instances based on existing data. In the context of information retrieval, data augmentation is typically applied to generate new queries, positive and negative examples. For example, the Inverse Cloze Task (ICT)~\cite{ICT} randomly selects a token span from a text segment to serve as the query, while the remaining tokens form the key. This is the opposite of the Cloze Task, where the remaining tokens are used as the query and the sampled token span serves as the candidate. This approach has been proven effective in pre-training\cite{PretrainingTasks, End-to-End}.

Additionally, the use of in-batch negatives is a method to expand negative examples. For a given query, the negatives are generated from the positive examples of other queries within the same batch. This method typically requires a larger batch size to generate more negatives for a query~\cite{SimpleFramework} and has been widely applied in open-domain retrieval scenarios~\cite{ICT, DPR, Unsupervised}.

Recently, researchers have also utilized LLMs to synthesize data for training embedding models. For instance, \citet{BGE-M3} used GPT-3.5 to generate questions for collected passages, while \citet{E5-mistral-7b-DBLP:journals/corr/abs-2401-00368} employed GPT-4 to first create task types and then construct queries, positive documents and hard negative documents based on these tasks. These models have set new state-of-the-art results on multiple benchmarks.


\section{Data Construction}
To address the challenges of asymmetric retrieval, queries in the training dataset should align with real-world user queries, which are often characterized by brevity and conciseness. As shown in Figure~\ref{fig:framework}, we propose an automatic method to generate queries based on case facts. We will introduce the details about the data generation in this section.

\begin{table*}[htbp]
    \centering
    \small
    \begin{tabular}{lrrrrr}
        \toprule
        Dataset & LeCaRD & CAIL2022-LCR & COLIEE2021 & COLIEE2022 &  \dataname{} \\
        \midrule
        Asymmetric          & \XSolidBold & \XSolidBold & \XSolidBold & \XSolidBold & \CheckmarkBold \\
        \# Query            & 107 & 40  & 900 & 1,198 & 100,060 \\
        Language            & Chinese & Chinese & English & English & Chinese \\
        \# Charge           &  20 & 19  & -- & -- &  210 \\
        Query Length        & 445 & 422 & 2,060 & 2,168 &  79 \\
        \bottomrule
    \end{tabular}
    \caption{Details of statistics of existing LCR datasets. The COLIEE dataset does not annotate the corresponding charges for the cases, so this table does not provide such information.}
    \label{tab:datasets}
    \vspace{-1em}
\end{table*}

\subsection{Query Generation} 
\label{sec: Query generation}
\textbf{Key Events Extraction.}
As all case documents are manually written by judges, there are many details and viewpoints contained in these documents, such as the names of every participant, their relationships, and the court discussion about each event. However, in real life, considering users' unfamiliarity with legal knowledge, the queries they search often only include key factual events.
To get the short queries as real-world user queries, we extract key information from the facts of legal cases gathered from online sources. Then, to do this efficiently, automatically, and at a large scale, our approach leverages a generative method based on open-source, large-scale language models. We employ an LLM to generate queries for our dataset. During the generation process, the model is first required to compress provided case facts into concise case descriptions, which only retain essential legal events. To guide the model, we furnish it with a task description and two illustrative examples within the prompt, ensuring effective and accurate query generation. 
The specific prompt is provided in appendix \ref{sec: Data Construction Details}.

\textbf{Anonymization.}
In the previous step, we also instruct LLM to remove entities such as personal names, locations, and dates from the cases. However, we found that approximately 30\% of cases still contain these entities, which are typically irrelevant to the key events and do not affect the final judgment. Besides, the shared entities between queries and candidates would provide a shortcut to the models, leading models trained on this data assign high relevance scores to the queries and candidates with the same entities and overlook critical legal events. Therefore, we implement a strategy to anonymize these entities.
Specifically, we utilize DeepTHULAC\footnote{https://github.com/thunlp/DeepTHULAC} for part-of-speech tagging of queries. Subsequently, specific information such as personal names, company names, locations, and time within the queries are replaced with semantically equivalent content. For instance, personal names are replaced with random usual names. 
This approach enables the model to better grasp the relationships between queries and key information, thereby enhancing the effectiveness of retrieval.

With the key events extraction and anonymization, we can generate a relevant query for every candidate case. The query-candidate pairs can serve as the training signals for LCR models.

\subsection{Knowledge-Driven Augmentation}
\label{正例增广}
Through the aforementioned method, we can construct large-scale query-candidate pairs that contain the same key facts. However, in real applications, we usually cannot find cases that are completely identical to the query. Therefore, to enable the model to handle a diverse range of queries in real-world scenarios, we further propose a knowledge-driven data augmentation method. 

Unlike open-domain information retrieval, in the LCR domain, it is not appropriate to judge whether two cases are similar based solely on the factual details of the case. The legal articles applicable to the case and the judgment results are also important~\cite{muser}.
Therefore, for a given query-candidate pair, we select the cases with similar legal articles and prison terms to the candidate as the augmented positive candidate. Specifically, we extract the main and ancillary legal articles from the ``Reason'' section of the case. Here, the main legal articles refer to those detailing specific charges, such as \textit{Article 133} from the Chinese Criminal Law, which defines and sets sentencing standards for the crime of traffic accidents. The ancillary legal articles refer to those outlining the impact of certain facts on sentencing, such as \textit{Article 67} from the Chinese Criminal Law, which defines self-surrender and its influence on the final sentencing. The content of these two articles is provided in appendix \ref{sec: law samples}. Additionally, we extract the charges and specific prison terms of the final judgment, such as death penalty and imprisonment, from the ``Judgment'' section. These extracted elements serve as the basis for positive augmentation.

Next, for each candidate case in the dataset, we identify a related case in which the main legal articles match those of the original candidate case, and the additional legal articles as well as prison terms are as similar as possible. This process results in a new positive example. This positive example is legally related to the original case, but because they are two completely different cases, it ensures that there is no overlap in the factual details. This process leads to a dataset that has been augmented with positive examples.

\subsection{Construction Details}
We collect 6.6 million criminal cases from China Judgment Online~\footnote{https://wenshu.court.gov.cn/}. Initially, we exclude criminal ruling documents (containing only content related to commutation) and retain only criminal judgment documents. Subsequently, we filter out cases with facts shorter than 100 Chinese characters, as the majority of criminal cases fall within this range. Using regular expressions, we match and extract information such as charges, legal articles, and judgments from the cases, eliminating those where such content couldn't be extracted via rules. In the end, there are about 2 million cases remained. From this pool, we randomly select 100 thousand cases to generate queries for each charge. Then, for each of these 100 thousand cases, we search for the most similar cases from the initial 2 million using charges, legal articles, and judgments as criteria, to augment new positive examples.

\begin{table*}[ht]
  \centering
  \small
  \begin{tabular}{l|c|cccccc}
    \toprule
    \multicolumn{1}{c|}{\multirow{2}{*}{Model}} & \multicolumn{1}{c|}{\multirow{2}{*}{Model Type}} & \multicolumn{6}{c}{LeCaRD} \\
     & & P@5 & P@10 & MAP & NDCG@10 & NDCG@20 & NDCG@30  \\
    \midrule
    BM25         & Traditional & 44.8 & 40.8 & 50.7 & 77.3 & 82.0 & 89.9 \\
    Chinese BERT & Pre-trained & 36.5 & 34.5 & 41.9 & 70.5 & 77.6 & 86.8 \\
    Lawformer    & Pre-trained & 40.6 & 38.5 & 45.6 & 74.4 & 80.0 & 88.5 \\
    SAILER       & Pre-trained & 51.8 & 46.5 & 59.7 & 86.0 & 89.5 & 93.9 \\
    ICT          & Augmentation & 37.6 & 36.7 & 45.6 & 72.2 & 78.9 & 87.5  \\
    CaseEncoder  & Augmentation & 50.8 & 45.8 & 57.7 & 83.6 & 87.4 & 92.7 \\
    BGE-M3 & Augmentation & 46.5  & 42.8 & 52.7 & 79.5 & 83.4 & 90.7 \\
    T\textsuperscript{2}Ranking 
                 & Fine-tuned & 43.7 & 40.0 & 49.3 & 75.6 & 81.6 & 88.9  \\
    GTE-Qwen1.5-7B-instruct  & Fine-tuned & 48.0 & 42.6 & 53.8 & 81.2 & 85.1 & 91.8  \\
    CAIL2022 Train & Fine-tuned & 46.7 & 44.1 & 52.7 & 80.5 & 85.0 & 91.0  \\
    \midrule
    Ours         & Augmentation & \textbf{56.3} & \textbf{49.6} & \textbf{63.5} & \textbf{87.3} & \textbf{89.9} & \textbf{94.5} \\
    \midrule
    \multicolumn{1}{c|}{\multirow{2}{*}{Model}} & \multicolumn{1}{c|}{\multirow{2}{*}{Type}} & \multicolumn{6}{c}{CAIL2022-LCR} \\
    & & P@5 & P@10 & MAP & NDCG@10 & NDCG@20 & NDCG@30 \\ \midrule
    BM25         & Traditional & 54.0 & 49.7 & 57.6 & 81.8 & 86.0 & 91.8 \\
    Chinese BERT & Pre-trained & 45.5 & 45.8 & 50.7 & 74.8 & 80.0 & 88.4 \\
    Lawformer    & Pre-trained & 53.0 & 50.5 & 57.5 & 84.5 & 87.9 & 93.0 \\
    SAILER       & Pre-trained & 60.5 & 54.2 & 65.7 & 91.9 & 94.3 & 97.0 \\
    ICT          & Augmentation & 51.0 & 47.7 & 53.5 & 81.5 & 85.2 & 91.5  \\
    CaseEncoder  & Augmentation & 58.0 & 54.2 & 63.6 & 91.7 & 93.6 & 96.5 \\
    BGE-M3 & Augmentation & 54.0 & 51.5 & 58.2 & 86.3 & 90.0 & 93.8 \\
    T\textsuperscript{2}Ranking 
                 & Fine-tuned & 54.5 & 52.2 & 59.3 & 86.6 & 89.4 & 94.1 \\
    GTE-Qwen1.5-7B-instruct  & Fine-tuned & 57.5 & 55.0 & 61.1 & 89.8 & 90.8 & 95.0  \\
    LeCaRD Train
                 & Fine-tuned & 56.0 & 53.5 & 59.6 & 88.6 & 91.5 & 94.7 \\
    \midrule
    Ours         & Augmentation & \textbf{65.0} & \textbf{58.0} & \textbf{67.7} & \textbf{94.0} & \textbf{94.7} & \textbf{97.4} \\
    \bottomrule
  \end{tabular}
  \caption{The main results of our model trained on \dataname{} and baseline models on LeCaRD and CAIL2022-LCR under the asymmetric retrieval setting.} 
  \label{tab: results}
\end{table*}

\subsection{Data Analysis}
With our method, we can easily construct the largest LCR dataset to date, with 100,060 query-case pairs, which is several hundred times larger than other LCR datasets available, and capable of supporting the training of existing data-hungry dense passage retrieval models. 

\paragraph{Diversity} 
In LCR, the diversity of charges can greatly benefit the performance of case retrieval. 
To achieve this, we carefully select cases across different charges and set a maximum threshold for the number of cases per charge. In total, we cover 210 charges, ensuring a diverse range of case descriptions.
Additionally, in our prompts for query generation, we include two examples to help the model learn how to generate queries. Since the generated queries are easily affected by examples, each time a query is generated, the examples in the prompt are randomly selected from a set of examples to ensure diversity in the generated queries.

It's worth noting that our data construction method is automated and doesn't rely on manual annotation. This makes it highly efficient for application to any criminal case with a clear structure. As a result, the dataset's size and coverage can be expanded rapidly, not limited solely to the numbers mentioned. In section \ref{sec: Civil}, we also apply the same method to generate data from civil cases. 

Due to the asymmetric nature of our dataset, the average query length is only 79 characters, which is more close to the real-world applications. 
Specific examples in the dataset can be found in Table \ref{tab: samples of datasets}, and we present the statistics of our constructed dataset and other widely-used LCR datasets in Table \ref{tab:datasets}.

\subsection{Model Training}
In this paper, we mainly focus on dense passage retrieval for legal cases. We adopt a dual-encoder architecture for all models. This involves separately encoding the query and the candidate cases to obtain query embeddings and candidate case embeddings and calculating the cosine similarity between them as the final similarity score. 

The training is conducted in an in-batch negative setting~\cite{DPR}.
In the in-batch negative setting, for each query in a batch with N training pairs, the negative examples are the positives of the other queries in the same batch, i.e., N-1 negative examples. However, when we use the newly identified positive examples from the dataset, some negatives may share the same charges, legal articles, or judgments with the positives, leading to false negatives that can impact the model training. To address this, during training, we set the cosine similarity between negatives with the same charges as the positive to $-\infty$. This is equivalent to removing these negatives from the negative set.

\section{Experiments}

\begin{figure*}
    \centering
    \begin{subfigure}[b]{0.24\textwidth}
    \centering
        \includegraphics[width=\textwidth]{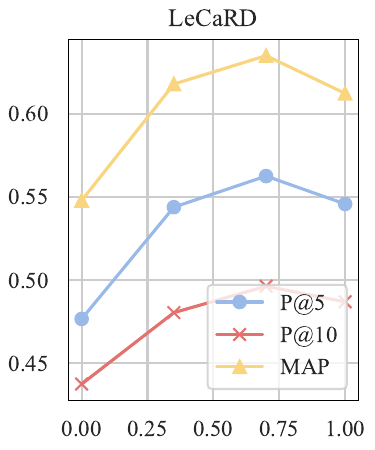}
        \caption{}
    \end{subfigure}
    \hfill
    \begin{subfigure}[b]{0.24\textwidth}
    \centering
        \includegraphics[width=\textwidth]{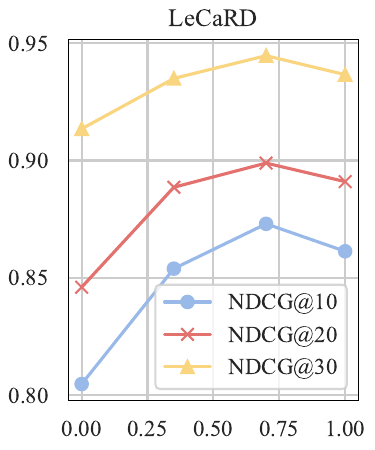}
        \caption{}
    \end{subfigure}
    \hfill
    \begin{subfigure}[b]{0.24\textwidth}
    \centering
        \includegraphics[width=\textwidth]{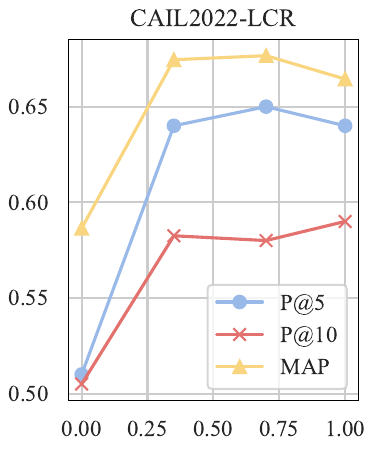}
        \caption{}
    \end{subfigure}
    \hfill
    \begin{subfigure}[b]{0.24\textwidth}
    \centering
        \includegraphics[width=\textwidth]{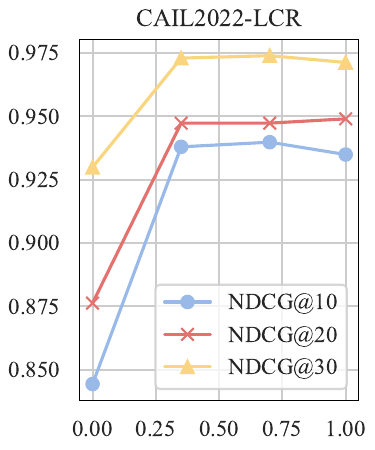}
        \caption{}
    \end{subfigure}
    \caption{Comparison of model performance with different proportions of augmented positive examples on LeCaRD and CAIL2022-LCR Datasets.}
    \label{fig:ablation-1}
\end{figure*}

\subsection{Datasets and Metrics} 
\label{sec: Datasets and Metrics}
In this paper, we focus on legal asymmetric retrieval, but existing datasets with human-annotated labels focus on symmetrical retrieval, where the queries are lengthy cases. Therefore, to better assess the model's performance in asymmetric retrieval, we adopt our method to simplify the query cases in benchmarks into a short version automatically.
To ensure the high quality of evaluation benchmarks, we manually check the generated queries, ensuring that the queries do not change the key events.
Specifically, we employ GPT-4 to generate the short version of queries and conduct quality testing by one of the authors. For case-to-case retrieval, we utilize the original datasets without query generation.



We adopt \dataname{} for training, and adopt two widely-used datasets for evaluation: (1)~\textbf{LeCaRD}~\cite{lecard} is a widely-used LCR evaluation dataset, which contains $107$ queries annotated by several legal practitioners.
(2)~\textbf{CAIL2022-LCR}~\footnote{http://cail.cipsc.org.cn/task3.html?raceID=3\&cail\_tag=2022} official testing set is furnished by the CAIL2022 organization, structured similarly to LeCaRD. We test our models on stage 2 of CAIL2022.
In both datasets, each query has $100$ candidate cases, but only $30$ of them are manually annotated. The annotations range from 0 (Both key facts and key circumstances are irrelevant) to 1 (Key facts are irrelevant but key circumstances are relevant), 2 (Key facts are relevant but key circumstances are irrelevant), and 3 (Both key facts and key circumstances are relevant). We only consider the annotated cases, and regard cases marked as 3 as relevant. 

As a retrieval task, we report normalized discounted cumulative gain (NDCG@10, NDCG@20, NDCG@30),  Precision (P@5, P@10), and Mean Average Precision (MAP). These evaluation metrics align with those used in LeCaRD, aiming to provide a comprehensive understanding of the model's performance across various aspects.

\subsection{Baselines} 
\label{sec: Baselines}
We compare our model with several competitive baselines, including:


\textit{Traditional Retrieval Model}: (1)~\textbf{BM25}~\cite{BM25} utilizes exact word matching to score documents based on their term frequencies and document lengths.

\textit{Pretrained Models}: 
    (1)~\textbf{Chinese BERT} is an adaptation of the original BERT model~\cite{BERT} for the Chinese. 
    (2)~\textbf{Lawformer}~\cite{Lawformer} is the first Chinese legal pre-trained model based on the longformer model~\cite{Longformer}.
    (3)~\textbf{SAILER}~\cite{SAILER} is a structure-aware pre-trained model for LCR, which employs an asymmetric encoder-decoder architecture for pre-training. 
    
\textit{Data Augmentation Method}: (1)~\textbf{Inverse Cloze task (ICT)}~\cite{ICT} is a data augmentation method for retriever pre-training, which randomly samples a span from a text segment as the query, while the remaining context as the candidate. (2)~\textbf{CaseEncoder}~\cite{CaseEncoder} constructs LCR data with fine-grained legal article information, which assumes that similar cases should contain similar legal articles. (3)~\textbf{BGE-M3}~\cite{BGE-M3} is trained on large-scale synthetic and labeled data, showing strong generalization performance.

\textit{Fine-Tuned Models}: (1)~\textbf{T\textsuperscript{2}Ranking}~\cite{T2Ranking} is a large-scale retrieval dataset in the open-domain. We directly utilize an open-source dual-encoder checkpoint, fine-tuned on the T\textsuperscript{2}Ranking dataset as our baseline model.
(2)~\textbf{GTE-Qwen1.5-7B-instruct}~\cite{DBLP:journals/corr/abs-2308-03281-gte} is based on a large language model of 7B parameters and harnesses a multi-stage contrastive learning, demonstrating broad applicability across various NLP tasks.
(3)~\textbf{LeCaRD / CAIL2022 Train} refers to the models trained with the instances contained in LeCaRD or CAIL2022. Details are provided in appendix \ref{sec: Experimental Details}. 
As one benchmark is used for training, we only present the results of this model on the other benchmark.

\subsection{Implementation Details} 
\label{sec: Implementation Details}


During evaluation, we employ a truncation strategy for lengthy candidates. Specifically, when the length of a candidate case exceeds the maximum sequence length of the utilized models, we truncate the case into multiple segments. Subsequently, we individually calculate the similarity score between each segment and the query, ultimately selecting the maximum similarity score as the final score for the candidate case.

The training batch size is set as 128 and the encoders are trained for up to 80 epochs with a learning rate of 1e-5 using Adam, linear scheduling with warm-up, and dropout rate 0.1. The maximum input sequence length was set to 2048. Additionally, our model reported in Table \ref{tab: results} utilizes positive augmentation data at a ratio of 70\%. That is, 30\% of the query-candidate pairs in the dataset consist of queries paired with their original cases, while the remaining 70\% of query-candidate pairs comprise simplified queries paired with cases newly identified using the method outlined in Section \ref{正例增广}. We randomly select 2048 samples from the dataset as the development set, with the rest used for training. 


\begin{table}[t]
  \centering
  \small
  \begin{tabular}{l|cccc}
    \toprule
    \multicolumn{1}{c|}{\multirow{2}{*}{}} & \multicolumn{4}{c}{LeCaRD} \\ 
    & P@5 & MAP & NDCG@10 & NDCG@30 \\
    \midrule
    Ours & \textbf{56.3} & \textbf{63.5} & \textbf{87.3} & \textbf{94.5} \\
    ~~w/o M & 52.0 & 58.0 & 84.1 & 92.8 \\
    \midrule \midrule
     & \multicolumn{4}{c}{CAIL2022-LCR} \\
     & P@5 & MAP & NDCG@10& NDCG@30 \\ \midrule
     Ours & \textbf{65.0} & \textbf{67.7} & \textbf{94.0} & \textbf{97.4} \\
     ~~w/o M & 59.5 & 63.4 & 90.4 & 96.1 \\
    \bottomrule
  \end{tabular}
  \caption{Comparison of model performance with and without false negative masking.}
  \label{tab: Mask}
\end{table}

\subsection{Main Result}
\label{sec: Experiment Result}

The overall results are presented in Table \ref{tab: results}. From the results, we can observe that:
(1)~Our model outperforms all baselines on both benchmarks by a large margin, achieving state-of-the-art performance. It indicates that using larger-scale and more comprehensive LCR data can greatly benefit task performance, which emphasizes the importance of developing data augmentation methods for LCR.
(2)~The traditional method, BM25, can outperform many models. Especially, BM25 can beat the models finetuned on T\textsuperscript{2}Ranking, which consisting millions of open-domain retrieval instances. It proves that LCR task is challenging and directly employing open-domain models can not achieve satisfactory results. That is because LCR requires the models to capture not only semantic relevance but also legal element relevance.
(3)~Compared to the pre-trained models, our model trained with \dataname{} can achieve siginificant performance improvements. The pre-training for LCR usually involves millions of cases and days of pre-training, which is computationally expensive. It shows the potential of scaling high-quality data for LCR, which can avoid expensive pre-training and yield superior performance. 
(4)~Our model can consistently outperform the data augmentation models and fine-tuned models. The existing data augmentation method can not generate high-quality data for LCR. Besides, existing open-domain data cannot benefit LCR performance, and the scale of existing manually annotated LCR datasets like LeCaRD cannot fulfill the requirements of training dense retrieval models, highlighting the importance of data scale rather than quality. Our proposed method to automatically construct large-scale data is effective in high-quality data generation.
We also extend the base model to LLM and train with our constructed data, as presented in appendix \ref{sec: scaling to LLM}.

\begin{table}[t]
  \centering
  \small
  \begin{tabular}{l|cccc}
    \toprule
    Models & BM25 & BERT & T\textsuperscript{2}Ranking & Ours \\
    \midrule
    Accuracy & 54.3 & 52.1 & 52.2 & \textbf{56.2} \\
    \bottomrule
  \end{tabular}
  \caption{The results on the CAIL2019-SCM dataset.} 
  \label{tab: civil}
  \vspace{-1.5em}
\end{table}

\begin{table*}[ht]
  \centering
  \small
  \begin{tabular}{l|c|cccccc}
    \toprule
    \multicolumn{1}{c|}{\multirow{2}{*}{Model}} & \multicolumn{1}{c|}{\multirow{2}{*}{Model Type}} & \multicolumn{6}{c}{CAIL2022-LCR} \\
    & & P@5 & P@10 & MAP & NDCG@10 & NDCG@20 & NDCG@30 \\ \midrule
    BM25         & Traditional & 50.5 & 49.8 & 55.1 & 80.2 & 82.7 & 90.5 \\
    Chinese BERT & Pre-trained & 46.5 & 47.0 & 52.6 & 78.2 & 81.8 & 89.9 \\
    Lawformer    & Pre-trained & 52.0 & 50.8 & 54.9 & 82.6 & 84.6 & 91.2 \\
    SAILER       & Pre-trained & 60.5 & 55.3 & 66.8 & 92.6 & 94.2 & 97.1 \\
    ICT & Augmentation & 48.5 & 47.0 & 52.2 & 79.6 & 82.9 & 90.6  \\
    CaseEncoder  & Augmentation & 63.5 & 56.0 & 65.6 & 92.8 & 94.1 & 96.9 \\
    BGE-M3 & Augmentation & 59.0 & 52.8 & 58.9 & 86.0 & 88.1 & 93.0 \\
    T\textsuperscript{2}Ranking 
                 & Fine-tuned & 56.5 & 50.8 & 57.4 & 83.4 & 86.7 & 92.2 \\
    GTE-Qwen1.5-7B-instruct  & Fine-tuned & 57.5 & 53.8 & 61.4 & 90.0 & 92.2 &  95.7 \\
    LeCaRD Train & Fine-tuned & 57.0 & 55.6 & 58.6 & 88.1 & 90.9 & 93.8 \\
    \midrule
    Ours         & Augmentation & \textbf{65.0} & \textbf{58.5} & \textbf{69.2} & \textbf{94.4} & \textbf{95.2} & \textbf{97.6} \\
    \bottomrule
  \end{tabular}
  \caption{The results of our model trained on LEAD and baseline models on CAIL2022-LCR under the traditional case-to-case symmetric retrieval setting.}
  \label{tab: traditional retrieval}
\end{table*}

\paragraph{Error Analysis} 
We have noticed the following errors: although our model is trained to handle short queries, it still struggles to identify the most relevant cases when the description of the case is only a few words long (e.g., \textit{``A killed B with a hammer then escaped.''}). At this point, among the most relevant cases, there are sometimes one or two cases with completely different charges (e.g., as hit-and-run). We assume that it's still difficult for the model to generate a vector representation of the legal elements contained in overly short queries.

Additionally, When two cases have many tokens in common, the model may overscore their similarity. For example, when retrieving medical malpractice cases, our model sometimes incorrectly ranks DUI (driving under the influence) cases highly because both types of cases often involve many concentration units (mg/mL).

\subsection{Ablation Study}
\label{sec: ablation}
We adopt a knowledge-driven data augmentation strategy for dataset construction. In this subsection, we conduct an ablation study to explore the impact of augmented positive examples.

\textbf{Proportion of Augmented Candidates.}
We adopt a knowledge-driven data augmentation strategy to make the query-candidate pairs with similar legal elements but diverse legal events. In this paragraph, to verify the effectiveness of the data augmentation, we conduct experiments with varying proportions of augmented positive examples within the dataset. Specifically, we present the results with the proportions as $\{0.00, 0.35, 0.700, 1.00\}$. The results are shown in Figure~\ref{fig:ablation-1}.

From the results, we can observe that: 
(1)~Compared with models without data augmentation ($0$\%), models trained with further data augmentation can achieve significant performance improvements for both two datasets and all metrics. It indicates that the knowledge-driven data augmentation methods can effectively match similar cases from the entire corpus and benefit the diversity of \dataname{}.
(2)~The optimal performance is achieved at 70\% and when the proportion reaches $100$\%, the model performance drops. This suggests that retaining a certain proportion of original cases as positive candidates is effective for LCR. We believe this is because these data instances help reduce the distance between simplified queries and original cases in the vector representation space, allowing the model to better comprehend the meaning of simplified queries in asymmetric retrieval scenarios.
Additionally, since the queries and the positive cases in this portion of the data come from the same cases, they have high semantic similarity, which also encourages the model to generate similar vector representations for semantically similar cases.

\textbf{False Negative Masking.} 
We adopt the in-batch negative sampling strategy to increase the scale of negative sampling. However, this training strategy will inevitably introduce false negative noises. To address this challenge, we adopt a false negative masking strategy, where the cosine similarity of negative candidates with the same charges is set to $-\infty$ during the training process. 
In this paragraph, we evaluate the effects of false negative masking strategy, with the results presented in Table \ref{tab: Mask}. We can find that removing the false negative masking strategy significantly deteriorates model performance on both datasets. This suggests that during the training process, many negative examples are indeed related to the query, and ignoring them can mitigate such interference.


\subsection{Civil Case Retrieval}
\label{sec: Civil}
Our method to automatically construct LCR datasets is flexible and can be easily extended to any case. Existing LCR works usually focus on criminal cases and overlook civil cases, which are more relevant to our daily lives. In this subsection, we construct a civil case retrieval dataset with the same construction method. Specifically, the judgment results of civil cases are more complex than criminal cases, and the knowledge-driven data augmentation strategy cannot be applied to civil cases. Therefore, here we present the results with no further candidate augmentation. Finally, we generate $77$k query-candidate pairs for civil cases. 
We utilize CAIL2019-SCM~\cite{CAIL2019-SCM} as the benchmark, which comprises 3036 triplets for the private lending cases, each consisting of three cases: A, B, and C. The task is to determine which of case, B or C, is more similar to A. We report the accuracy of several models that are not limited to criminal cases, and our model in Table \ref{tab: civil}.
Despite using only simplified queries and their corresponding original cases as training data, our model can achieve the best performance on this test set. This demonstrates that simple asymmetric retrieval data can also enable the model to understand legal elements, validating the robustness of our approach.

\subsection{Case-to-Case Symmetric Retrieval}
In this paper, we mainly focus on asymmetric LCR and our large-scale dataset can also benefit the traditional case-to-case symmetric retrieval setting. In this subsection, we evaluate the models in the traditional setting. The results are shown in Table~\ref{tab: traditional retrieval}. From the results, we can observe that (1)~Our model still outperforms other models by a large margin, indicating that our constructed asymmetric retrieval dataset is not only effective for asymmetric retrieval tasks but also performs excellently in traditional case retrieval scenarios. This suggests that our model effectively learns to identify similar legal elements through augmented positive examples. (2)~The baseline models can achieve superior performance on the asymmetric retrieval setting. That is because the lengthy query can provide more detailed information for models to retrieve similar cases. The short queries require the models to associate the key events and legal knowledge to capture relevance between the query and candidates, which presents a great challenge for existing models. Therefore, we encourage the community to devote more efforts to asymmetric LCR.


\section{Conclusion}
In this paper, we propose a method for automatically constructing high-quality, asymmetric legal case retrieval datasets. 
We construct the largest LCR dataset to date, with over one hundred thousand query-candidate pairs, surpassing existing datasets by a hundredfold. 
We conduct experiments on two widely-used datasets, achieving state-of-the-art performance in LCR tasks. Moreover, our method is highly versatile, showing superior performance in civil case retrieval as well.

\section*{Limitations}
In this paper, we discuss the limitations of this paper:
(1)~We construct a large-scale synthetic LCR dataset for Chinese cases. Our method is language-agnostic and can also be applied to cases in other countries, which is worth exploring in the future.
(2)~We only fine-tune our model with LCR synthetic data. In the future, we can combine it with open-domain synthetic data to train an embedding model capable of multi-task applications. 

\section*{Acknowledgement}
This work is supported by the National Science and Technology Major Project (2022ZD0160502), the National Natural Science Foundation of China (62106126), the National Social Science Fund of China (21AZD143), the Guoqiang Institute, Tsinghua University, Tsinghua-Toyota Joint Research Fund, Beijing Advanced Innovation Center for Future Blockchain and Privacy Computing.
\bibliography{anthology,custom}

\newpage

\begin{table}[t]
\centering
\small
\begin{tcolorbox}[colback=white,boxrule=0.5mm, arc=4mm]

\textbf{System:} 

As a legal expert, you are capable of extracting key elements from the basic information of a case.

\textbf{User:} 

I have a dataset for searching cases by case. However, the basic information of the cases in my dataset is too long. I will send you the basic information of these cases, please help me simplify them, and greatly shorten their length while retaining key legal elements. You can remove non-key names, locations, etc., but do not delete important elements for case judgments.

\textbf{Assistant:} 

I understand your requirement.

\textbf{User:} 

In August 2015, XX Co., Ltd. obtained the right to use state-owned construction land in XX Lake area of XX District through public auction, and developed the "XX" project. The defendants Alice, Bob, Charlie, David, and Edward are villagers of XX Village, XX Street Office, XX District, within the scope of the project. The demolition and compensation within the scope of the land has been implemented by the Management Committee of XX in 2014. After the "XX" project started construction on July 13, 2016, Alice, Bob, and others proposed to the construction party to contract part of the project such as excavation of retaining walls and earthwork excavation of foundation piles. Because they did not have the construction qualification, the project developer did not agree. From July 20 to August 12, 2016, Alice, Bob, Charlie, David, and Edward convened at "XX" and "XX" Tea House to discuss obstructing the construction, and raised 70,000 yuan for the obstruction fund. At the same time, Bob, Charlie, David, and others used WeChat groups and phone calls to invite and mobilize more than 20 villagers from their village and their family members to obstruct the construction of the "XX" construction site by methods such as locking the gate, pulling the power switch, insulting, and standing on construction machinery, causing the construction site to be unable to proceed normally. The economic loss caused by the obstruction during the construction period was appraised by XX District Price Certification Center as 124,530 yuan.

\textbf{Assistant:} 

A company obtained the right to use construction land through auction and compensation has been implemented, but villagers nearby conspired to obstruct construction, raising a fund of 70,000 yuan and mobilizing more than 20 villagers to repeatedly obstruct the construction site, causing an economic loss appraised at 124,530 yuan.






\end{tcolorbox}
\caption{Prompt and an example for generating short queries, where specific information such as names and locations have been anonymized.}
\label{tab: prompt for GPT-4}
\end{table}

\appendix
\section{Appendix}
\label{sec:appendix}

\subsection{Data Construction Details}
\label{sec: Data Construction Details}

To generate concise case descriptions from case facts, we employ a large-scale generative language model, for query generation. The input instructions and a sample case description, along with its original case fact, are shown in Table \ref{tab: prompt for GPT-4}.

The generated case description retains all the legal elements from the original case fact while omitting the rest of the content. The original case fact, being part of a court judgment, contains a plethora of details to comprehensively describe the case's proceedings. However, including these details as part of a real-world user query is redundant.

\begin{table}[t]
\centering
\small
\begin{tcolorbox}[colback=white,boxrule=0.5mm, arc=4mm]

\textbf{Query Case 1:} 

The defendant, on a certain day while passing through a neighborhood,  \textcolor{blue}{stole} a motorcycle worth approximately 3,000 yuan. Subsequently, the defendant rode the motorcycle to another city, intending to \textcolor{red}{sell} it to someone, but was apprehended on the spot by the owner. The stolen motorcycle has been recovered and returned to the victim. The defendant has confessed to their crime.

\textbf{Positive Case 1:} 

The People’s Procuratorate of Hanshan County accuses: On the evening of September 24, 2017, the defendant Li Jun walked to the entrance of the old transportation bureau dormitory lane opposite Hanshan No. 2 Middle School, and \textcolor{blue}{stole} the Jixiangshi brand two-wheeled electric bike parked there by reconnecting the electric wire. The next evening, the defendant Li Jun rode the stolen electric bike to the Shanghai Qiqiang Electric Bike Shop located at Wangmei Road in Hanshan County \textcolor{red}{for sale}. Since the price negotiation with the shop owner was not successful, he then hid the electric bike under the building of Han City River and River Water Conservancy Construction and Installation Co., Ltd. The appraisal price of the stolen electric bike was 1760 yuan. On October 1, 2017, the defendant Li Jun was arrested at his home in Motang Village, Chengbei Administrative Village, Huanfeng Town, Hanshan County by the Hanshan County Public Security Bureau. On October 2, 2017, the Hanshan County Public Security Bureau returned the stolen vehicle to the victim Mao.

\textbf{Query Case 2:} 

Defendant Alice was driving a car while \textcolor{red}{intoxicated}, \textcolor{blue}{rear-ending} another vehicle and causing property damage. Alice was determined to be fully responsible. Alice's \textcolor{red}{blood alcohol content exceeded the legal limit}.

\textbf{Positive Case 2:} 

After investigation, it was found that on January 20, 2012, at around 8:10 PM, the defendant, Yu, had dinner and drank alcohol with friends. \textcolor{red}{After drinking}, he drove the vehicle with license plate Shaanxi AWB062 home. While driving north along Mingguang Road and approaching the intersection with Fengcheng 8th Road, he failed to brake in time and collided with the rear \textcolor{blue}{end} of the vehicle with license plate Shaanxi AFU210, driven by Guo Guangcheng, who was waiting at the traffic light. This caused Guo's vehicle to rear-end the vehicle in front, with license plate Shaanxi A05V90, driven by Zhao Ming, resulting in a traffic accident involving damage to all three vehicles. The public security authorities apprehended the defendant, Yu, at the scene. The road traffic accident report determined that the defendant, Yu, was fully responsible for the accident, while Guo Guangcheng and Zhao Ming bore no responsibility. It was determined that the defendant, Yu, had a \textcolor{red}{blood alcohol concentration of 180.51 mg/100 ml}. Further investigation revealed that on February 2, 2012, the defendant, Yu, paid Zhao Ming 12,000 yuan for vehicle repairs. On February 10, 2012, the defendant compensated Guo Guangcheng 65,000 yuan, after which Guo Guangcheng transferred ownership of the vehicle with license plate Shaanxi AFU210 to Yu.

\end{tcolorbox}
\caption{Two examples of data constructed using our method. The similar legal key elements in the cases are marked with the same color.}
\label{tab: samples of datasets}
\end{table}

\subsection{Experimental Details}
\label{sec: Experimental Details}

\paragraph{Training with LeCaRD}
LeCaRD training set annotates 30 cases for relevance to each query. When constructing the dataset, for each query $Q_i$, all cases with a relevance score of 3 are designated as $\{P_{i1}, P_{i2}, ..., P_{in}\}$, while the remaining cases are designated as $\{N_{i1}, N_{i2}, ..., N_{im}\}$. If $m < n$, then $m-n$ cases are randomly selected from the $70$ unannotated cases to form $\{N_{i(m+1)}, N_{i(m+2)}, ..., N_{in}\}$. Each training datum consists of one query, one positive case, and one negative case, denoted as $(Q_i, P_{ij}, N_{ij})$, where $i=1,2,...,107$ and $j=1,2,...,n$. This process results in a training set of size 1,112. The remaining implementation details are same as those described in Section \ref{sec: Implementation Details}. Existing datasets usually contain limited annotated pairs and cannot fulfill the requirements for the training of data-hungry neural models.

\subsection{Addition Experiment Result}
\label{sec: Addition Experiment Result}

We also conducted experiments on the original LeCaRD dataset under the traditional case-to-case symmetric retrieval setting, and the results are shown in Table \ref{tab: traditional retrieval on LeCaRD}. 
Here, we present the results of all baseline models and the models trained on \dataname{} with different proportions of augmented positive examples. 

From the results, we can observe that similar to the results on the CAIL2022-LCR dataset, our dataset, \dataname{} can significantly benefit the performance of traditional case-to-case symmetric retrieval.

\begin{table*}[ht]
  \centering
  \small
  \begin{tabular}{l|c|cccccc}
    \toprule
    \multicolumn{1}{c|}{\multirow{2}{*}{Model}} & \multicolumn{1}{c|}{\multirow{2}{*}{Model Type}} & \multicolumn{6}{c}{LeCaRD} \\
    & & P@5 & P@10 & MAP & NDCG@10 & NDCG@20 & NDCG@30 \\ \midrule
    BM25         & Traditional & 40.7 & 39.5 & 48.9 & 73.5 & 78.8 & 87.7 \\
    Chinese BERT & Pre-trained & 36.8 & 36.0 & 42.8 & 70.2 & 77.0 & 86.5 \\
    Lawformer    & Pre-trained & 40.2 & 37.7 & 46.7 & 73.6 & 79.7 & 88.3 \\
    SAILER       & Pre-trained & 49.5 & 44.3 & 57.7 & 84.7 & \textbf{88.9} & \textbf{93.7} \\
    ICT & Augmentation & 36.3 & 35.6 & 45.1 & 70.0 & 77.0 & 86.6 \\
    CaseEncoder  & Augmentation & 49.2 & 45.8 & 57.2 & 83.5 & 87.5 & 92.9 \\
    BGE-M3 & Augmentation & 45.6 & 41.4 & 51.8 & 77.2 & 81.9 & 89.9 \\
    T\textsuperscript{2}Ranking 
                 & Fine-tuned & 43.9 & 40.1 & 49.9 & 75.7 & 81.1 & 89.0 \\
     GTE-Qwen1.5-7B-instruct  & Fine-tuned & 45.6 & 40.7 & 51.3 & 77.9 & 83.0 & 90.4  \\
    \midrule
    Ours  (0\%)       & Augmentation & 45.0 & 42.0 & 51.7 & 77.8 & 82.8 & 90.1 \\
    Ours  (35\%)      & Augmentation & 51.8 & 46.4 & 59.0 & 83.1 & 87.2 & 92.5 \\
    Ours  (70\%)      & Augmentation & \textbf{54.4} & 47.1 & 60.9 & 84.3 & 87.8 & 93.0 \\
    Ours  (100\%)     & Augmentation & 52.3 & \textbf{47.3} & \textbf{61.8} & \textbf{84.7} & 88.2 & 93.3 \\
    \bottomrule
  \end{tabular}
  \caption{The results of our model trained on LEAD and baseline models on LeCaRD under the traditional case-to-case symmetric retrieval setting.}
  \label{tab: traditional retrieval on LeCaRD}
  \vspace{-1em}
\end{table*}

\subsection{Scaling to LLM}
\label{sec: scaling to LLM}
We also scaled our base model to LLM and then fine-tuned it using our data.
\subsubsection{Implementation Details}
LLM is typically trained on the Next Token Prediction task, utilizing causal attention and Last Token Pooling strategy. To adapt the model into an Embedding Model, we first modified it to bidirectional attention and Mean Pooling strategy.

We employed the open-source generative language model MiniCPM~\cite{DBLP:journals/corr/abs-2404-06395-minicpm}. 
For our training setup, we set the batch size to 128 and trained the model for up to 10 epochs with a learning rate of 1e-4 using Adam, linear scheduling. The softmax score was set to 0.2. Due to computational constraints, we limited the sequence length to 512 and employed LoRA~\cite{DBLP:conf/iclr/HuSWALWWC22-lora} with a rank of 16. Additionally, we enabled mixed precision training with bfloat16. 
We did not use the false negative masking strategy here.

\subsubsection{Main result}
As shown in Table \ref{tab: model scaling results}, although MiniCPM is a generative language model, the results of training it directly with LCR data still significantly surpass the strongest baseline, SAILER. This demonstrates the powerful potential of scaling models in LCR. By incorporating data from other domains, we can train large models that perform exceptionally well across multiple tasks.
\begin{table*}[ht]
  \centering
  \small
  \begin{tabular}{l|cccccc}
    \toprule
    \multicolumn{1}{c|}{\multirow{2}{*}{Model}} & \multicolumn{6}{c}{LeCaRD} \\
    & P@5 & P@10 & MAP & NDCG@10 & NDCG@20 & NDCG@30 \\ \midrule
        SAILER  & 51.8 & 46.5 & 59.7 & 86.0 & 89.5 & 93.9 \\
        Ours (MiniCPM) & \textbf{53.8} & \textbf{47.8} & \textbf{62.3} & \textbf{87.4} & \textbf{90.3} & \textbf{94.8} \\ 
    \bottomrule 
  \end{tabular}
  \caption{The results of our model based on MiniCPM, trained on \dataname{}, under the asymmetric retrieval setting on LeCaRD}
  \label{tab: model scaling results}
  \vspace{-1em}
\end{table*}
\subsection{Articles of the \textit{Criminal Law of the People's Republic of China}}
\label{sec: law samples}
\textit{\textbf{Article 67 }}

\textit{[General Voluntary Surrender] If, after committing a crime, the offender voluntarily surrenders and truthfully confesses their crime, it is considered voluntary surrender. For offenders who voluntarily surrender, a lighter or mitigated punishment may be imposed. If the crime is minor, the punishment may be waived.}

\textit{[Special Voluntary Surrender] If a criminal suspect, defendant, or convict under compulsory measures truthfully confesses to other crimes not yet known to the judicial authorities, it is considered voluntary surrender.}

\textit{Even if a criminal suspect does not meet the conditions for voluntary surrender specified in the previous two paragraphs, a truthful confession of their crime can lead to a lighter punishment; if the truthful confession prevents particularly severe consequences, a mitigated punishment may be imposed.}

\textit{\textbf{Article 133}} 

\textit{[Traffic Accident Crime] Violating traffic and transportation regulations resulting in a major accident that causes serious injury, death, or significant property damage shall be punished by imprisonment of up to three years or criminal detention. If the offender flees the scene of the accident or if there are other particularly egregious circumstances, the punishment shall be imprisonment of three to seven years. If fleeing the scene results in a person's death, the punishment shall be imprisonment of seven years or more.}

\textit{Article 133-1 [Dangerous Driving Crime] Driving a motor vehicle on the road under any of the following circumstances shall be punished by criminal detention and a fine:}

\textit{(1) Racing in a particularly egregious manner;}

\textit{(2) Driving a motor vehicle while intoxicated;}

\textit{(3) Seriously exceeding the passenger limit or the speed limit while engaged in school bus or passenger transport services;}

\textit{(4) Violating safety management regulations for the transport of hazardous chemicals, thereby endangering public safety.}

\textit{If the owner or manager of the motor vehicle is directly responsible for the actions specified in items (3) and (4) of the preceding paragraph, they shall be punished according to the preceding paragraph.}

\textit{If the actions specified in the preceding two paragraphs also constitute other crimes, the more severe punishment shall apply.}

\textit{Article 133-2 [Obstructing Safe Driving Crime] Using violence against or forcibly taking control of the operating equipment of the driver of a public transportation vehicle in operation, thereby interfering with the normal operation of the vehicle and endangering public safety, shall be punished by imprisonment of up to one year, criminal detention, or control, and may also be fined.}

\textit{If the driver of a public transportation vehicle in operation abandons their post, fights with others, or assaults others, thereby endangering public safety, they shall be punished according to the preceding paragraph.}

\end{document}